\documentclass[aps,prl,floatfix,12pt,superscriptaddress,longbibliography]{revtex4-1}

\usepackage{verbatim}
\usepackage{graphicx}
\usepackage{color}
\usepackage{amsmath}
\usepackage{amssymb}
\usepackage{braket}
\usepackage{subfig}
\usepackage[hidelinks]{hyperref}

\begin{document}
	
	\title{Optimization strategies for modulation transfer spectroscopy applied to laser stabilization}
	
	\author{Tilman Preuschoff}
	\author{Malte Schlosser}
	\author{Gerhard Birkl}
	\affiliation{Institut f\"{u}r Angewandte Physik, Technische Universit\"{a}t Darmstadt, Schlossgartenstra\ss e 7, 64289 Darmstadt, Germany}

	%\email{\authormark{*}apqpub@physik.tu-darmstadt.de} %% email address is required
	
	\homepage{www.iap.tu-darmstadt.de/apq} %% author's URL, if desired
	\date{\today}

\begin{abstract}
We present a general analysis for determining the optimal modulation parameters for the modulation transfer spectroscopy scheme. The results are universally valid and can be applied to spectroscopy of any atomic species requiring only the knowledge of the effective linewidth $\Gamma_{eff}$. A signal with optimized slope and amplitude is predicted for a large modulation index $M$ and a modulation frequency comparable to the natural linewidth of the spectroscopic transition. As a result of competing practical considerations, a modulation index in the range of $3 \le M \le 10$ has been identified as optimal. This parameter regime is experimentally accessible with a setup based on an acousto-optic modulator. An optimized signal for spectroscopy of the rubidium D2 line is presented. The signal shape and the dependence on the modulation parameters are in very good agreement with the theoretical description given. An experimental procedure for achieving a strong suppression of residual amplitude modulation is presented. Based on the optimized signal, we demonstrate long-term laser stabilization resulting in a laser linewidth of 150\,kHz (16\,s average) and a frequency stability of 18\,kHz (rms) over 15 hours.\\
		\\
		\bfseries{Original Citation (Open Access):\\
			Optics Express 26, 24010-24019 (2018); \href{https://doi.org/10.1364/OE.26.024010}{DOI: 10.1364/OE.26.024010}} 
\end{abstract} 
\maketitle

\section{Introduction}

Stable laser sources with narrow linewidth and high long-term stability are of great importance for experiments in many fields of physics. Especially, applications that require resonant excitation of an atomic transition demand a long-term stability and a linewidth well below the natural linewidth of the transition. Suitable sources are realized by active frequency stabilization to atomic references with high resolution Sub-Doppler spectroscopy \cite{Demtroeder2014}. A large bandwidth of the stabilization loop is crucial to obtain narrow linewidths. It is a particularly challenging task to implement a stabilization scheme with a sufficient long-term stability and a large bandwidth at the same time. Sub-Doppler spectroscopy schemes that produce DC-type signals, such as polarization spectroscopy (PS), can be realized with a large signal bandwidth but they are inherently susceptible to long-term drifts. Frequency modulation spectroscopy schemes provide an improved long-term stability \cite{Demtroeder2015}.   Among them, the modulation transfer spectroscopy (MTS) scheme is significant due to the coherent nonlinear four-wave mixing transfer process being efficient for modulation frequencies in the order of the natural linewidth \cite{Ducloy1981,Ducloy1982,Camy1982,Shirley1982}. Compared to frequency modulation spectroscopy schemes based on the incoherent process of modulated spectral hole burning (MSHB) \cite{Negnevitsky2013},  this large modulation frequency features a large bandwidth and a reduced $1/f$ noise. The absolute accuracy is enhanced by a strong suppression of features in the signal from non-closed atomic transitions and the Doppler background.

In this work, we present a species-independent theoretical analysis of the MTS technique for identification of the optimal parameter space. This is complemented by experimental results on spectroscopy and laser stabilization on the ${}^{85}$Rb D2 line. In Section~\ref{sec:theory}, a theoretical description of the MTS signal is presented. Based on the model discussed in \cite{Ducloy1981,Ducloy1982,Camy1982,Shirley1982} we give an analytical expression of the MTS signal under the assumption of an ideal two-level atom. We find that this analytical model is fully sufficient and more suitable to identify the optimal modulation parameter regime than a full multi-level treatment where only numerical solutions are available \cite{Noh2011,Noh2015}. Signal slopes and amplitudes are evaluated extending the previous work for small modulation indices \cite{Jaatinen1995,Carron2008} to the regime of large indices. A setup based on an acousto-optic modulator (AOM) described in Section~\ref{sec:setup} is used to access the optimal parameter regime experimentally. The setup is optimized to produce a low-noise stabilization signal with a bandwidth of 100\,kHz. The long-term stability is enhanced by introducing an experimental alignment procedure which strongly reduces residual amplitude modulations. As a result, an active feedback for compensation of amplitude modulation effects \cite{Zhang2014} is not necessary.  The obtained line shape is compared to the theoretical description in Section~\ref{sec:line}. The predicted signal slopes and amplitudes as a function of the modulation index are verified for different modulation frequencies. In Section~\ref{sec:performance}, the linewidth and the long-term stability of an MTS-stabilized laser system are presented.  

\section{Theory: line shape and optimal modulation parameters}\label{sec:theory}

Consider a frequency-modulated pump field with amplitude $E_0$, carrier frequency $\omega$, modulation index $M$, and modulation frequency $\omega_m$:

\begin{equation}
E_{pump}(t) = E_0\, \cos(\omega t + M \sin(\omega_m t)).
\end{equation}
Fourier decomposition yields $E_{pump}$ as a sum of carrier and sidebands: 

\begin{equation}\label{eqn:F-Reihe}
E_{pump}(t) = E_0 \sum\limits_{k \in \mathbb{Z}} J_k(M) \cos\left((\omega + k \omega_m) t \right).
\end{equation}
The amplitudes are given by  Bessel functions of the first kind $J_k$. Note that in general, sidebands of order $k \gg M$ are negligible. The pump field is superimposed with a counter-propagating probe field $E_{probe}(t) = E_{0,p} \sin \omega t$ in a medium with a resonance at $\omega_0$. Taking the three fields, $E_{probe}$, the carrier of $E_{pump}$, and one sideband of $E_{pump}$ at $\omega + k \omega_m$, the nonlinear process of degenerate four-wave mixing (DFWM) creates a fourth field with components at frequencies $\omega \pm k \omega_m$ propagating in direction of $E_{probe}$ for $\omega$ close to $\omega_0$ \cite{Raj1980}. Considering all sidebands of $E_{pump}$, DFWM leads to a complete coherent modulation transfer from $E_{pump}$ to the resulting field in direction of $E_{probe}$. Under the assumption of a two-level system for the medium and in the low absorption limit, the power spectrum of the resulting field can be calculated analytically \cite{Ducloy1981,Ducloy1982}. Adjacent sidebands in the spectrum give a beat signal at $\omega_m$ \cite{Camy1982}. Demodulation with a reference signal proportional to $\cos(\omega_m t + \phi)$ yields the MTS signal

\begin{multline}\label{eqn:MTS_Signal}
	S_{M,\omega_m, \phi}(\Delta)   =  \frac{C}{\sqrt{\omega_m^2 +\Gamma_{eff}^2}} \sum\limits_{k \in \mathbb{Z}} J_k(M) J_{k-1}(M)  \\
\times\left[\left(L_{\frac{k-2}{2},\omega_m}(\Delta) + L_{\frac{k+1}{2},\omega_m}(\Delta)\right) \cos \phi  \right. 
+ \left. \left(D_{\frac{k-2}{2},\omega_m}(\Delta) - D_{\frac{k+1}{2},\omega_m}(\Delta)\right) \sin \phi   \right]
\end{multline}
as a function of the detuning  from the atomic transition $\Delta$ = $\omega - \omega_0$, with the components 

\begin{equation}
L_{n,\omega_m}(\Delta) = \frac{1}{1+\left(\frac{\Delta - n \omega_m}{\Gamma_{eff}}\right)^2}
\end{equation}
and

\begin{equation}
 D_{n,\omega_m}(\Delta) = \frac{ \frac{\Delta - n \omega_m}{\Gamma_{eff}}}{1+\left(\frac{\Delta - n \omega_m}{\Gamma_{eff}}\right)^2}
\end{equation}
as introduced in \cite{Shirley1982}. In this expression, $\Gamma_{eff}$ is the effective linewidth of the atomic transition, given by the natural linewidth $\Gamma$ and additional broadening effects \cite{Eble2007}.  By pairwise comparison of the terms in $k$ and $-k+1$, it is straightforward to see that $S_{M,\omega_m,\phi}$ is always an odd function in $\Delta$ and $S_{M,\omega_m,\phi}(0) = 0$. This makes it a suitable choice for a spectroscopic signal with a well defined center frequency applicable to laser stabilization. We evaluate Eq.~(\ref{eqn:MTS_Signal}) in 25 orders of $J_k$ which is a valid approximation for $M \le 20$. As a measure of the signal quality, the slope  at $\Delta = 0$ and the peak-to-peak amplitude are studied as a function of the experimentally accessible modulation parameters $\omega_m$ and $M$. Equation~(\ref{eqn:MTS_Signal}) can be rewritten in the form

\begin{equation}
S_{M, \omega_m, \phi} (\Delta) = A_{M, \omega_m}(\Delta) \cos \phi + B_{M, \omega_m}(\Delta) \sin \phi
\end{equation}
with the in-phase part $A_{M, \omega_m}$ and the quadrature part $B_{M, \omega_m}$. It is apparent that there exists an optimal phase $\phi_{s}$ with

\begin{equation}
\tan \phi_s = \left. \frac{\frac{\mathrm{d}}{\mathrm{d}\Delta} B_{M, \omega_m} }{\frac{\mathrm{d}}{\mathrm{d}\Delta} A_{M, \omega_m}}\right|_{\Delta = 0}
\end{equation}
that maximizes the signal slope $\mathrm{d}S/\mathrm{d}\Delta$ at $\Delta=0$ as a function of phase $\phi$ for fixed parameters $M$ and $\omega_m$. However, since the functions $A_{M,\omega_m}$ and $B_{M,\omega_m}$ are not extremal for the same $\Delta$, it is not straightforward to give an analytical expression for the phase $\phi_a$ that maximizes the amplitude. This phase is obtained numerically by maximizing the amplitude of $S_{M,\omega_m,\phi}$ over $\Delta$ and $\phi$.

\begin{figure}[ht!]
	\centering
	\subfloat[\quad Normalized signal slope at $\Delta = 0$]{\includegraphics[height= 0.38 \textwidth]{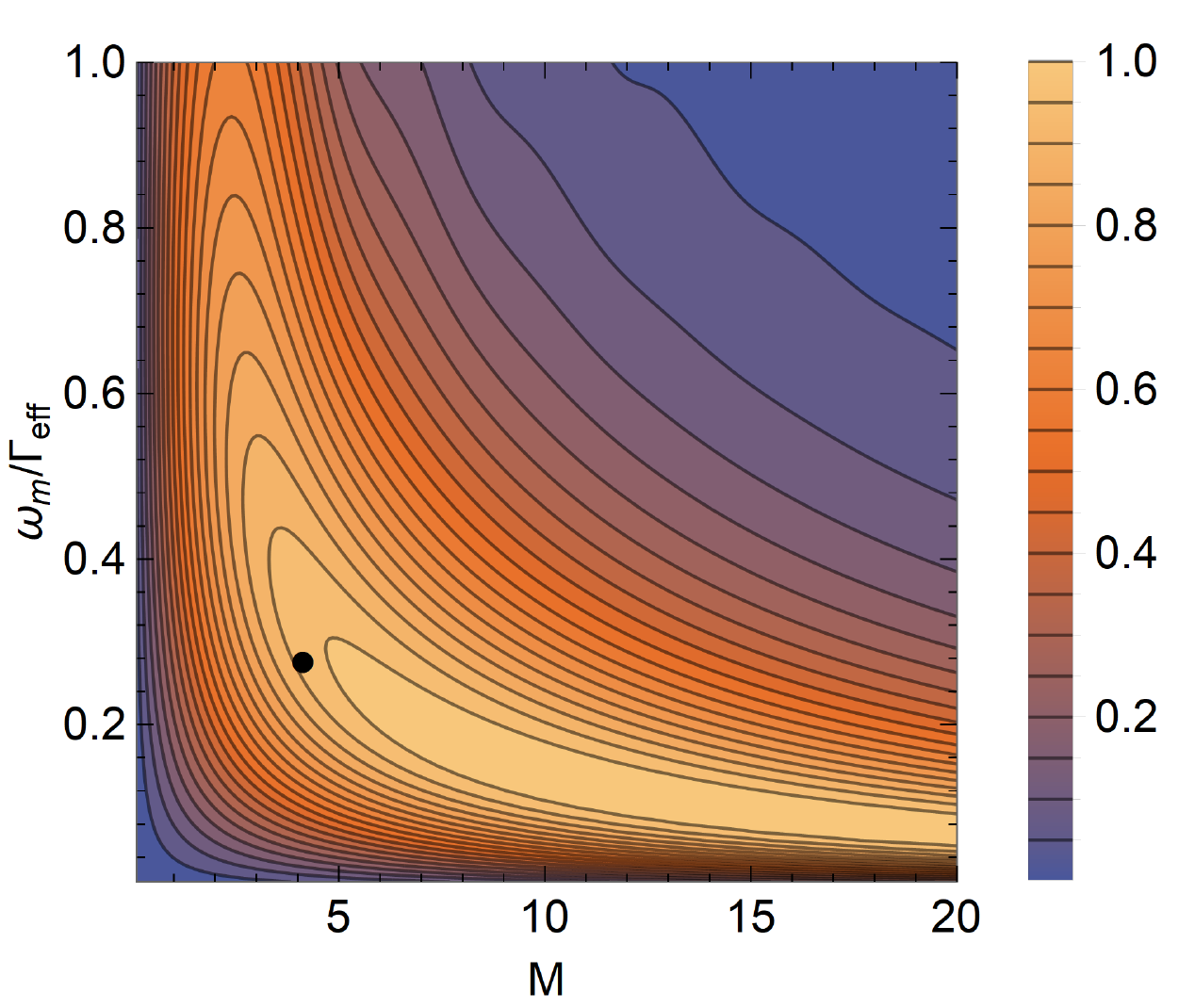}} \quad
	\subfloat[\quad Phase $\phi_s$ for maximal slope]{\includegraphics[height= 0.38 \textwidth]{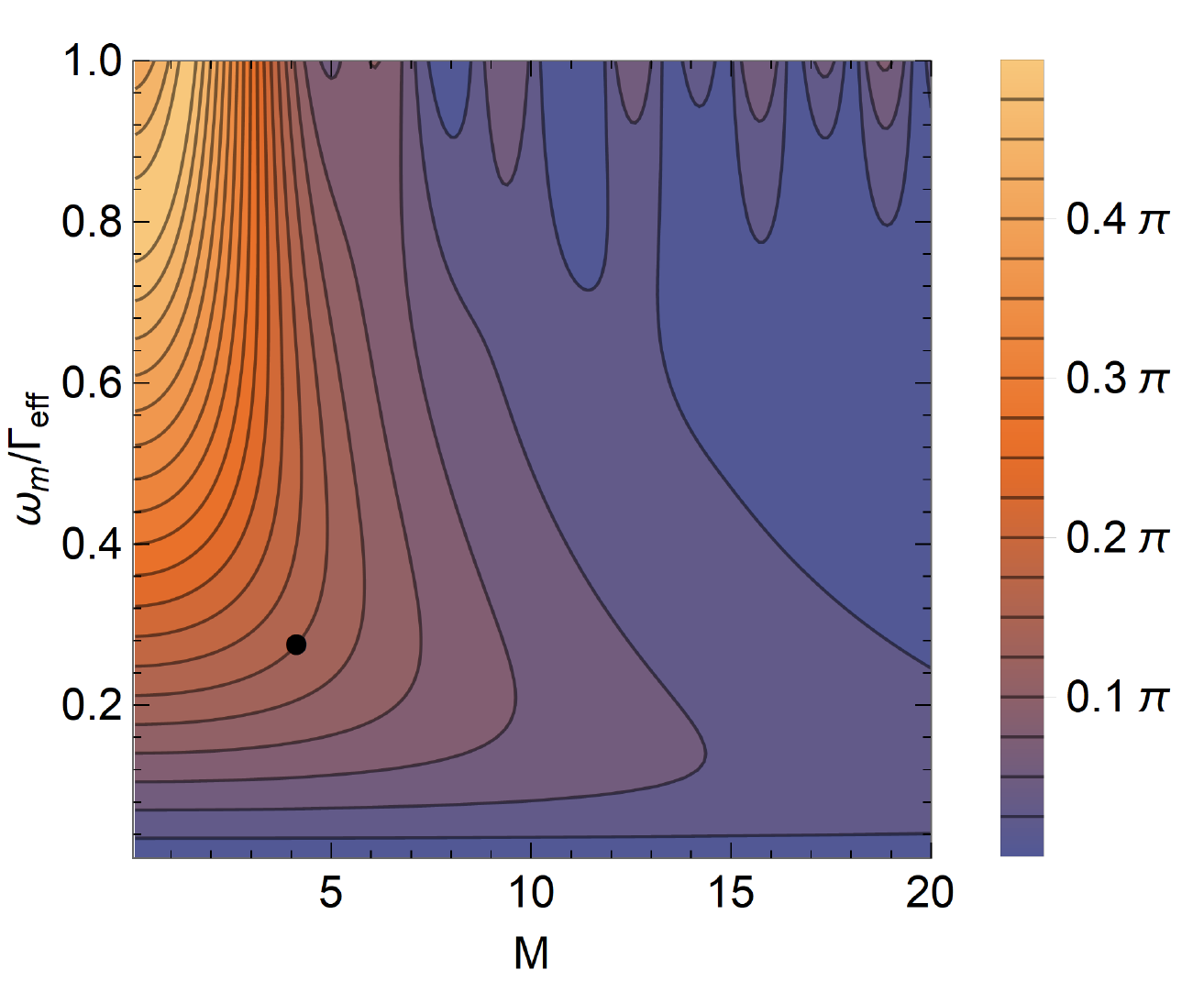}}\\
	\subfloat[\quad Normalized signal amplitude]{\includegraphics[height = 0.38 \textwidth]{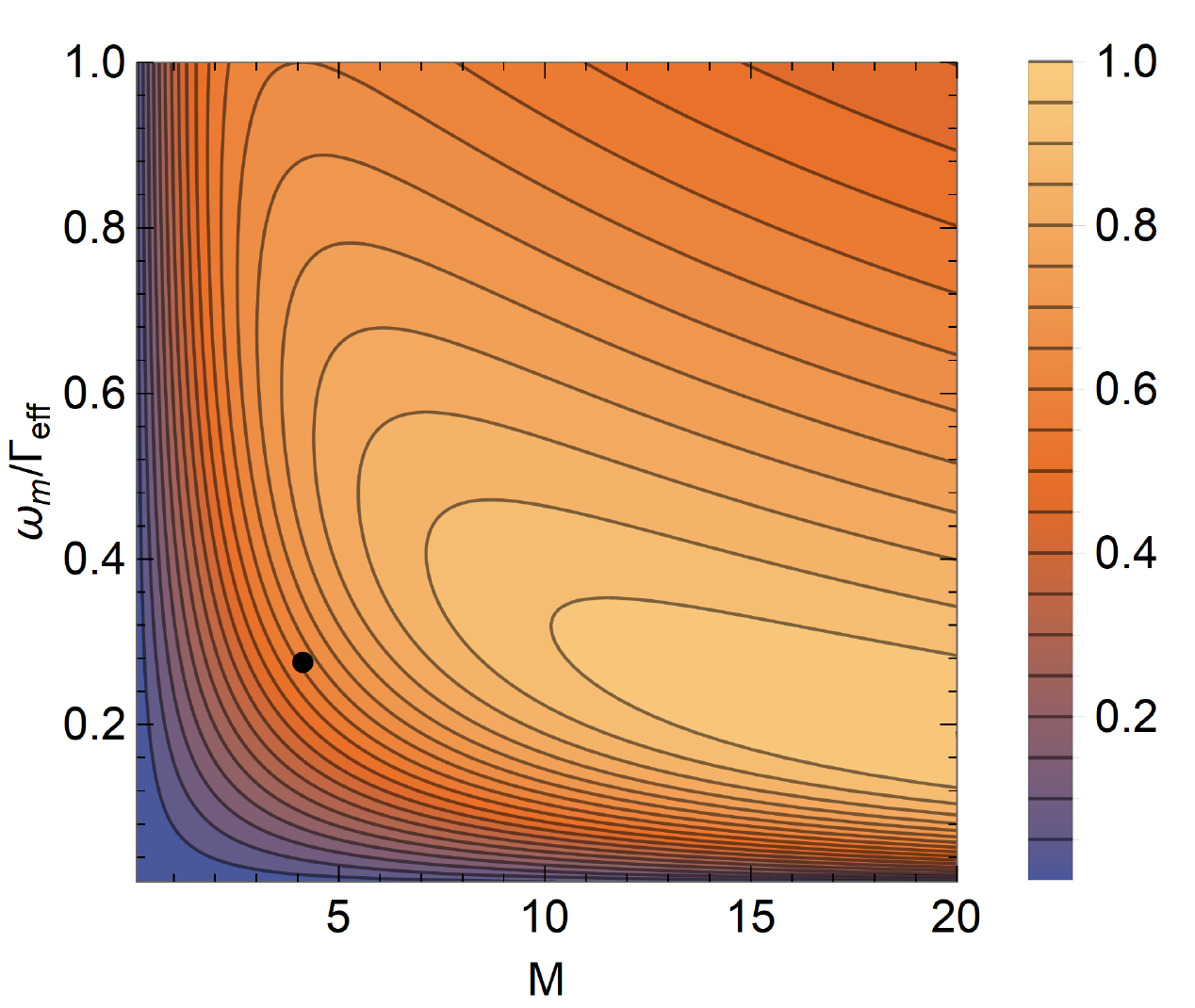}} \quad
	\subfloat[\quad Phase $\phi_a$ for maximal amplitude]{\includegraphics[height = 0.38 \textwidth]{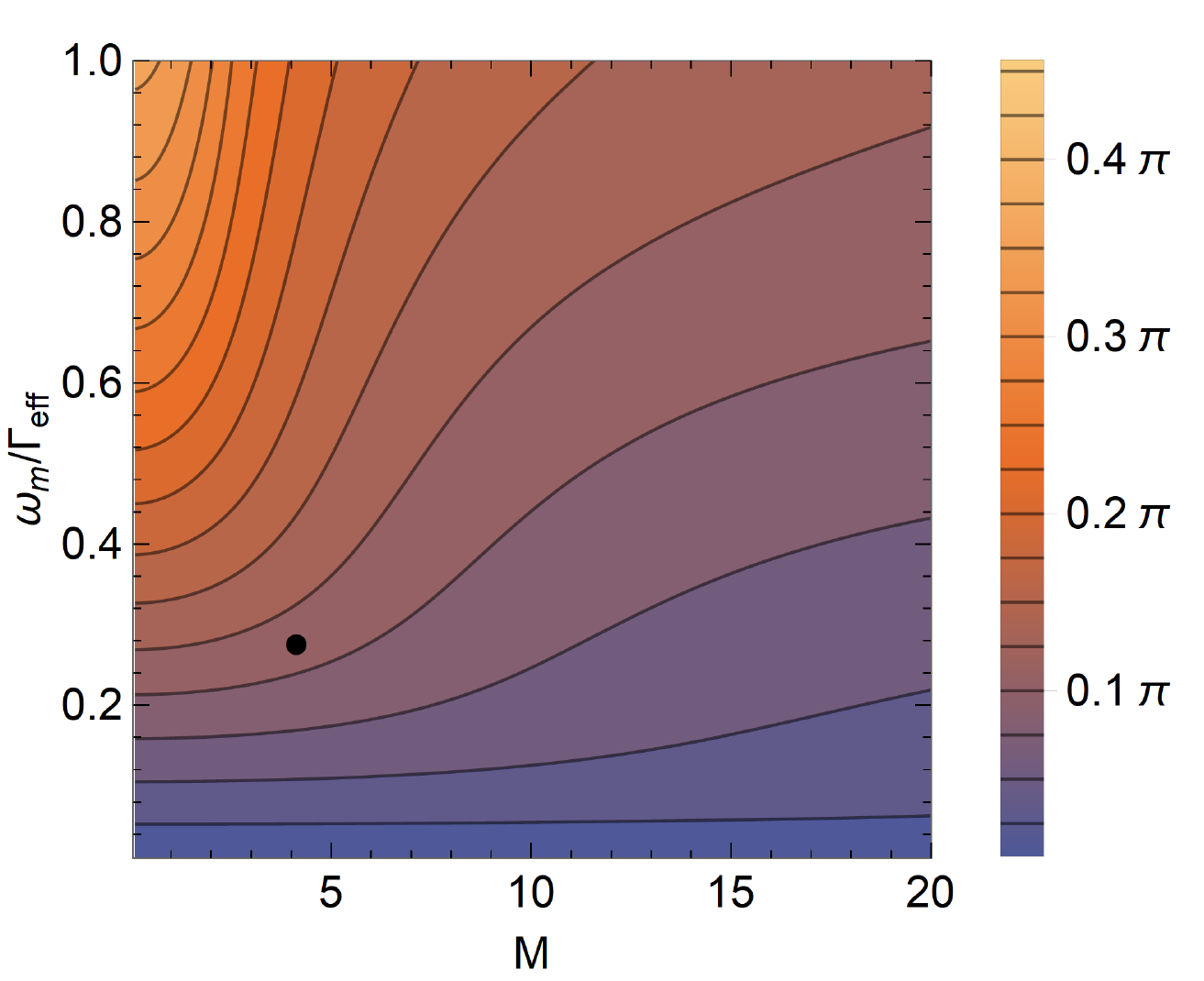}}
	\caption{\label{fig:ContPl}Signal slope (a) and amplitude (c) as a function of modulation index $M$ and modulation frequency $\omega_m$ in units of the linewidth $\Gamma_{eff}$. Both are normalized to the maximal values appearing in the plot. Amplitudes are calculated for the optimal demodulation phase $\phi = \phi_a$ and slopes for $\phi = \phi_s$. The optimal phases $\phi_s$ and $\phi_a$ are shown in (b) and (d). The black dots mark the experimental parameters discussed in Section~\ref{sec:line}. \label{fig:phase}}
\end{figure}

Figures~\ref{fig:ContPl}(a) and \ref{fig:ContPl}(c) show signal slope and amplitude as a function of $M$ and $\omega_m$ in units of $\Gamma_{eff}$ for $\phi_a$ and $\phi_s$ respectively. In general, a larger modulation index yields a larger slope and amplitude while the optimal value for $\omega_m$ is shifted to lower frequencies. Slopes and amplitudes are normalized to the maximal values in the plot obtained for $M = 20$. The achievable slope and amplitude as a function of $M$ saturate well below this value. We performed additional calculations up to $M = 100$ which show that the maximum slope increases only by 0.3\% and the maximum amplitude only by 2\% beyond $M = 20$. The absolute value of the optimal modulation frequency $\omega_m$ is proportional to $\Gamma_{eff}$ which is in general not identical to the natural linewidth $\Gamma$. The optimal phases $\phi_s$ and $\phi_a$ are shown in Figs.~\ref{fig:phase}(b) and \ref{fig:phase}(d) as a function of $\omega_m$ and $M$. Slope and amplitude show a similar behavior: For a large $M$ and a small $\omega_m$, a larger fraction of the in-phase part is optimal ($\phi_s,\phi_a \approx 0$); for a small $M$ and a large $\omega_m$, a larger fraction of the quadrature part is optimal ($\phi_s,\phi_a \approx \pi/2$). Globally, slope and amplitude are maximal in the case of large $M$ and small $\omega_m$ where the optimal phases are very similar as well ($\phi_s \approx \phi_a$). Although it is not possible to maximize slope and amplitude simultaneously, there exist modulation parameters that yield 85\% of the maximal slope and of the maximal amplitude simultaneously. Those can only be found for $M > 5$.

Note, that defining the quality of the signal by slope and amplitude alone neglects the occurrence of deteriorating structures in the spectroscopy signal which have been observed experimentally and theoretically in the case of $M < 1$ \cite{Carron2008}. Those structures can be avoided if the DFWM doublets in Eq.~(\ref{eqn:MTS_Signal}) are not resolved. This is the case for $\omega_m / \Gamma_{eff} < 4/ (3\sqrt{3}) \approx 0.77$ for the in-phase part and $\omega_m / \Gamma_{eff} < 4/ \sqrt{3} \approx 2.31$ for the quadrature part. As can be seen in Fig.~\ref{fig:ContPl}, optimal values for $\omega_m/\Gamma_{eff}$ are below those bounds for $M > 3$ where an unperturbed spectroscopic reference can be generated. On the other hand, small values of $\omega_m/\Gamma_{eff}$ are experimentally unfavorable for spectroscopy due the occurrence of additional features  \cite{Negnevitsky2013} resulting from MSHB and for laser stabilization due to the reduction in servo bandwidth. As a consequence, modulation indices in the range of $3 \le M \le 10$ emerge as an excellent choice for optimized performance. 

\section{Experimental setup}\label{sec:setup}

\begin{figure}[ht!]
	\centering
	\scriptsize
	\def\svgwidth{0.75 \columnwidth}
	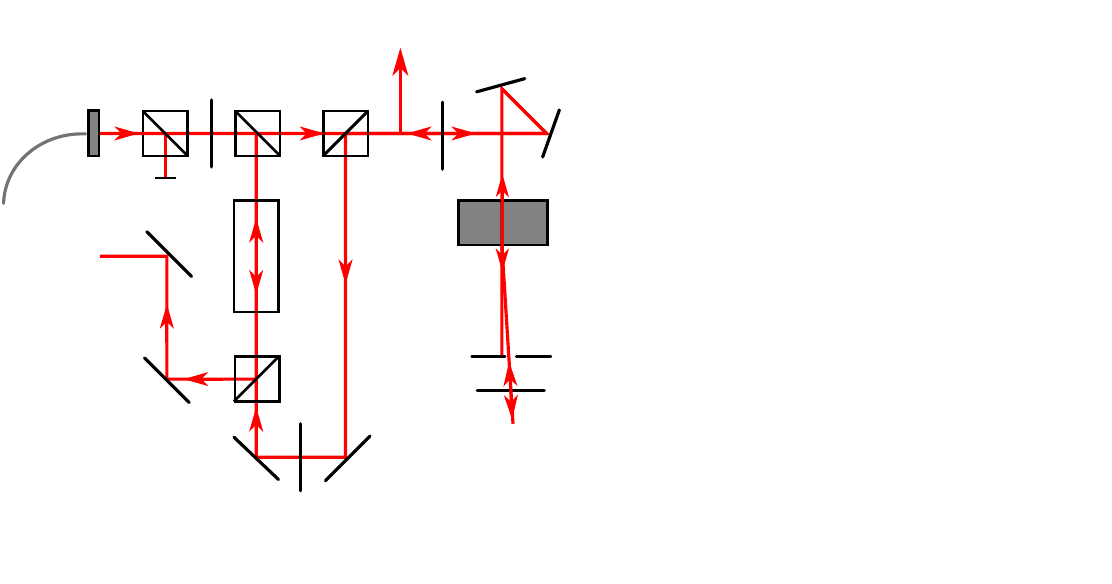
	\caption{\label{fig:setup}AOM-based MTS setup (see text for details). Designation of optical elements: PBS for polarizing beam splitter, PD for photodiode, L for focusing lens. PBS1 serves for cleaning the input beam polarization. Optical beam paths are colored in red and electronic signal lines in blue.}
\end{figure}

The MTS setup shown in Fig.~\ref{fig:setup} is based on an AOM (Crystal Technology 3080-120) in double pass configuration. This modulator allows for a large modulation index and a large range of modulation frequencies. The carrier frequency of the pump beam is shifted by $2 \omega_c$ = $2 \pi \cdot 160\,$MHz relative to the probe beam which shifts the center frequency of the spectroscopic signal by $\omega_c$ and avoids unwanted signal features due to reflections, e.g. at the rubidium glass cell. The cell (TRIAD Technology TT-RB-75-V-P) is enclosed in a mu-metal box which reduced the magnetic field to a value below $1\,\mu$T and is heated to $(315\pm 1)\,$K for an optimal signal \cite{Bing2014}. Pump and probe beams are superimposed using polarizing beam splitters (PBS) which allows for an exact and lossless overlay with optimal lin$\perp$lin polarization configuration \cite{Sun2016}. The probe beam intensity is detected with a home-built low-noise photodiode (PD1). For the probe beam 150\,$\mu$W and for the pump beam 400\,$\mu$W of optical power are used. Accounting for all occurring losses, a total power of 1.5\,mW at the fiber output port is needed for the spectroscopy. Probe and pump beams have a beam waist of approximately 0.5\,mm. The radio-frequency signal driving the AOM is produced by a signal generator (Agilent E4421B) that is frequency-modulated externally by a DDS synthesizer (Novatech 409B) at modulation frequencies up to $\omega_m = 2\pi\cdot 2.5\,$MHz. A modulation depth of $\Delta\omega = M \omega_m$ up to $2\pi \cdot 10.35\,$MHz can be obtained by half of the modulation depth applied to the AOM due to the double pass configuration. The photodiode signal is amplified (Mini-Circuits ZFL500-LN) and demodulated by mixing (Mini-Circuits ZRPD-1) with a reference signal obtained from a second channel of the DDS synthesizer. This configuration allows a precise and stable control of the demodulation phase $\phi$. However, it is not possible (and also not necessary) to determine the absolute value of $\phi$ since it is strongly depended on the beam position in the AOM. The relevant relative phase $\phi$ between the two DDS output signals was optimized experimentally for every chosen set of modulation parameters. To obtain a suitable stabilization signal, the frequency mixer output is filtered by a home-built second order Butterworth filter with a cutoff frequency of 100\,kHz. The maximal obtainable bandwidth is limited to a fraction of $\omega_m$ given by the filter type. For larger bandwidths, a higher order filter can be used \cite{Negnevitsky2013}.

\begin{figure}[ht!]
	\centering
	\includegraphics[width = 0.5 \textwidth]{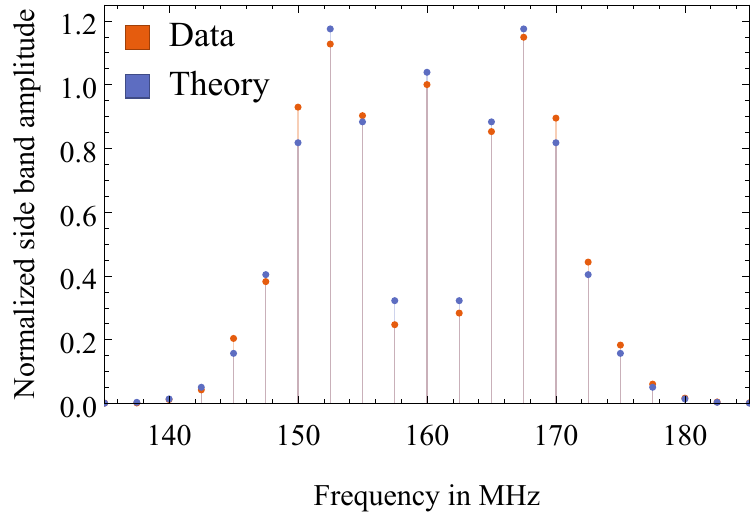} 
	\caption{\label{fig:pump_spec}Beat spectrum between pump and probe beam for a modulation depth $\Delta\omega = 2\pi \cdot 10.35\,$MHz at $\omega_m = 2\pi \cdot 2.5\,$MHz  compared to a calculated ideal spectrum with $M = 4.14$. }
\end{figure}

A well studied issue in MTS setups is residual amplitude modulation (RAM) \cite{Jaatinen2008,Jaatinen2009}. This effect renders the sideband structure asymmetrically and leads to an intensity depended shift of the spectroscopic line center. In the present setup, RAM is reduced through careful alignment of the beams in the AOM and in the spectroscopy cell: in a separate measurement, the beat spectrum between pump and probe beam is observed directly (Fig.~\ref{fig:pump_spec}). The spectrum is used to adjust the AOM double pass such that asymmetries in the modulation spectrum are almost eliminated. This significantly reduces RAM induced by the AOM. Additional RAM can be induced by an imperfect spatial overlap of probe and pump beam in the medium  \cite{Jaatinen2008}. The beam overlap is optimized by minimizing the remaining asymmetries in the MTS signal. This strategy allows us to compensate both mechanisms separately and make the signal shape independent of intensity variations to a high degree (see Section~\ref{sec:performance}). If necessary, PD3 and an AOM on the input side of the optical fiber (not shown) can be used to stabilize the overall power in the setup to a sub 1\%-level in order to reduce the effect of remaining RAM even further. 

\section{Results: spectroscopic signal and dependence on modulation parameters}\label{sec:line}

\begin{figure}[ht!]
	\centering
	\includegraphics[width = 0.5 \textwidth]{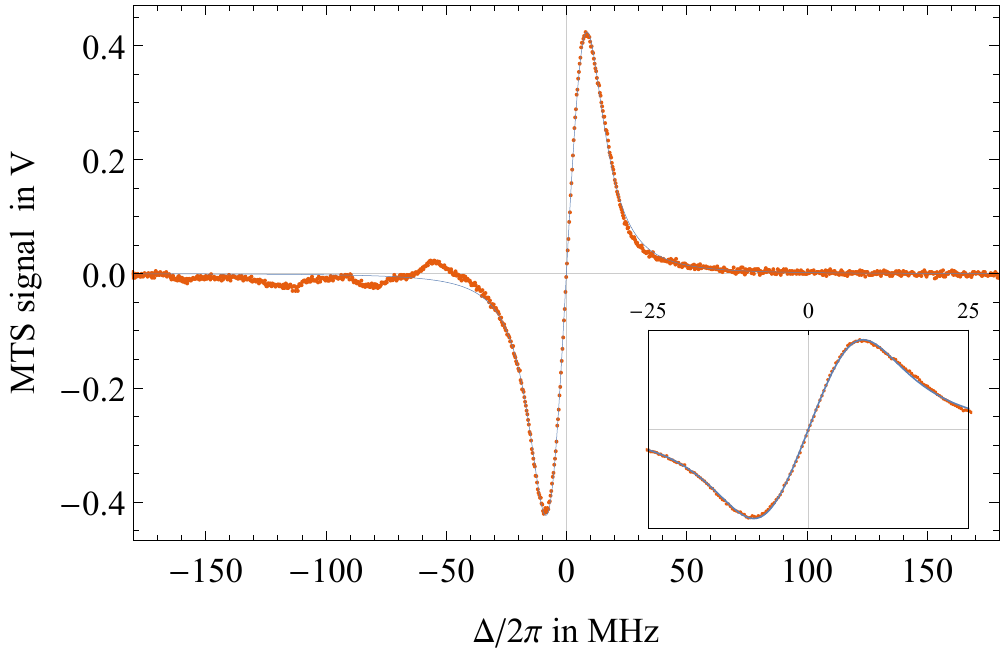}
	\caption{Experimental MTS signal of the dominant ${}^{85}$Rb hyperfine transition $\ket{5^2S_{1/2}, F = 3, m_F} \rightarrow \ket{5^2P_{3/2}, F' = 4, m'_F}$ (orange). The theoretical model of Eq.~(\ref{eqn:MTS_Signal}) is fitted to the data yielding $\Gamma_{eff} = 2\pi\cdot (9.03 \pm 0.42)\,$MHz (blue line). The frequency scale has been centered to the  zero-crossing of the signal. \label{fig:ErrorSig}}
\end{figure}

The MTS setup is implemented for spectroscopy on the ${}^{85}$Rb D2 line with  a home-built interference-filter stabilized external cavity diode laser system (ECDL) \cite{Baillard2006,Martin2016}. Scanning the laser frequency over the atomic resonance gives the MTS signal shown in Fig.~\ref{fig:ErrorSig} with the closed hyperfine transition $\ket{5^2S_{1/2}, F = 3, m_F} \rightarrow \ket{5^2P_{3/2}, F' = 4, m'_F}$ as the dominant feature. The frequency scale is derived from the Doppler-free absorption signal recorded by PD2 (see Fig.~\ref{fig:setup}). As expected, the MTS signal features a flat baseline and a strong suppression of the nearby transitions. Visible are small features from the crossover resonances $F = 3 \rightarrow F' = 2,4$ and $F = 3 \rightarrow F' = 3,4$ that become more significant for low $\omega_m$. This behavior has been observed previously \cite{Negnevitsky2013}. In Fig.~\ref{fig:ErrorSig}, the MTS signal is compared to the theoretical prediction of Eq.~(\ref{eqn:MTS_Signal}). The obtained signal shape is in very good agreement with theory (see also inset). The signal features power broadening \cite{Eble2007}. This has been accounted for by fitting $\Gamma_{eff}$ yielding $\Gamma_{eff} = 2\pi\cdot (9.03 \pm 0.42)\,$MHz $= (1.49 \pm 0.07) \,\Gamma$. Putting an emphasis on a maximal slope and a high modulation frequency, we selected a modulation frequency of $\omega_m = 2\pi \cdot 2.5\,$MHz and a modulation index of $M = 4.14$ for this measurement.  The corresponding operating point with $\omega_m/\Gamma_{eff} = 0.277 \pm 0.015$ is depicted as black dot in Fig.~\ref{fig:ContPl}. For the chosen parameters, according to Fig.~\ref{fig:ContPl}(a,c) we achieve 90\% of the maximal signal slope and 60\% of the maximal signal amplitude.

\begin{figure}[ht!]
	\centering
	\subfloat{\includegraphics[width= 0.48 \textwidth]{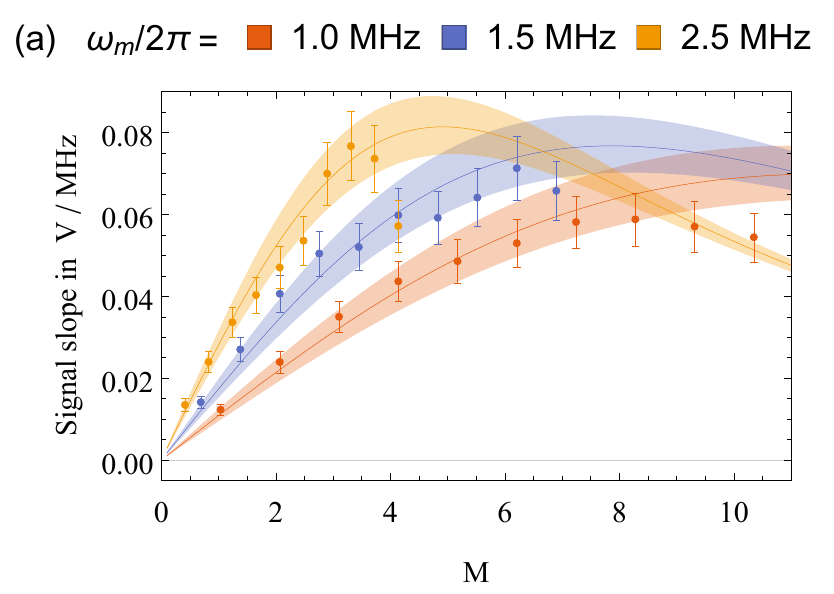}}\quad
	\subfloat{\includegraphics[width = 0.48 \textwidth]{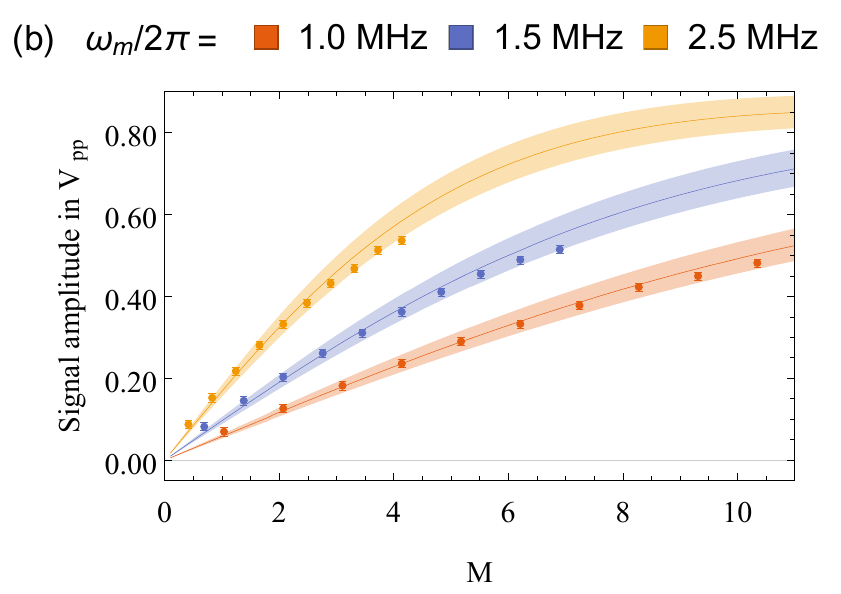}}
	\caption{Experimentally obtained signal slopes (a) and amplitudes (b)  as a function of $M$ for different $\omega_m$ compared to the theoretically predicted values (lines). The shaded areas mark the uncertainty of the theoretical prediction due to uncertainties in the determination of $\Gamma_{eff}$. \label{fig:AmplSl} }
\end{figure}

In Fig.~\ref{fig:AmplSl}, measured signal slopes and amplitudes are shown as a function of $M$ by varying $\Delta\omega$ for three different values of $\omega_m$. For every $\omega_m$, the phase $\phi$ has been optimized experimentally at $\Delta\omega = 2\pi\cdot 5.18\,$MHz for maximum amplitude. The theoretical curves have been obtained by calculating signal slopes and amplitudes for a constant phase $\phi_a$ determined for $M = 2\pi\cdot 5.18\,$MHz$/\omega_m$ and $\omega_m$ according to Fig.~\ref{fig:phase}(d). For every $\omega_m$, the scaling factor $C$ in Eq.~(\ref{eqn:MTS_Signal}) has been determined by a fit to the amplitude data set. In comparison to Fig.~\ref{fig:ErrorSig}, the measurements were performed with a slightly lower total light power, resulting in a lower effective linewidth $\Gamma_{eff} = 2\pi\cdot (8.17 \pm 0.38)\,$MHz. The shaded areas mark the uncertainty of the theoretical prediction due to uncertainties in the determination of $\Gamma_{eff}$. Figure~\ref{fig:AmplSl} shows good agreement between experimental results and the theoretically predicted behavior for modulation indices of up to $M = 10$. The deviation for large $M$ is caused by the limited bandwidth of the AOM which induces a modified pump spectrum not being accounted for by our model. Thus, the preferable parameter range of large modulation indices becomes experimentally accessible with AOM-based modulation transfer spectroscopy. This extends previous work on MTS with electro optic modulators (EOM) \cite{Zi2017,Sun2016,Bing2014,Xiang-Hui2009,Carron2008,Eble2007} and AOMs \cite{Negnevitsky2013,Martinez2015}.
 
\section{Laser stabilization: laser linewidth and long-term stability}\label{sec:performance}

\begin{figure}[ht!]
	\centering
	\subfloat[\label{fig:beat}]{\includegraphics[height = 0.31 \textwidth]{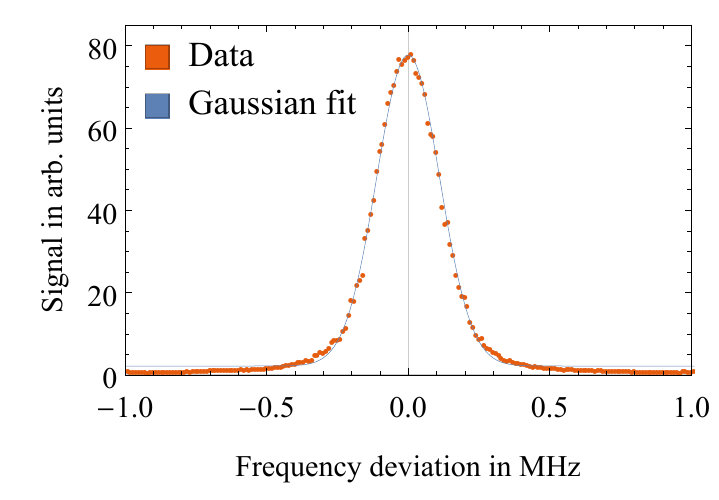}}\quad
	\subfloat[\label{fig:longterm}]{\includegraphics[height = 0.31 \textwidth]{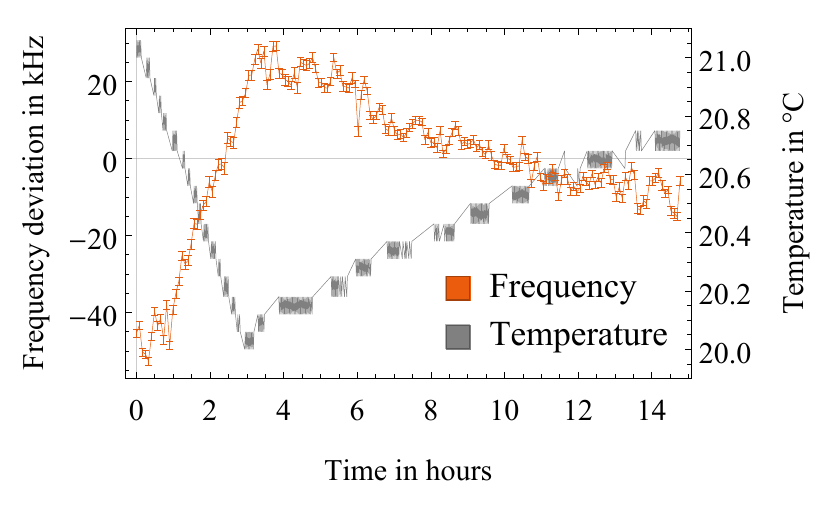}}
	\caption{(a) Beat note spectrum between the MTS-stabilized ECDL system and a PS-stabilized ECDL system. The Gaussian fit yields a combined linewidth of 260\,kHz. The laser linewidth of the MTS-stabilized system is $(150\pm 7)\,$kHz. (b) Center frequency of the beat note spectrum of an MTS-stabilized ECDL and MSHB-stabilized ECDL observed over a period of 15~hours. Both laser systems stayed in lock over the full duration. The variation in beat frequency tracks the change in ambient temperature.\label{fig:stabi}}
\end{figure}

The optimized MTS signal presented in Fig.~\ref{fig:ErrorSig} is used to lock an ECDL system to the ${}^{85}$Rb transition $\ket{5^2S_{1/2}, F = 3, m_F} \rightarrow \ket{5^2P_{3/2}, F' = 4, m'_F}$. The MTS-stabilized ECDL is compared to an ECDL system of the same type stabilized via polarization spectroscopy (PS) and an ECDL system stabilized via frequency modulation spectroscopy based on modulated spectral hole burning (MSHB). Locking is performed by a home-built PI controller that acts on the effective cavity length with a piezoelectric actuator in a low bandwidth loop ($\approx 1$\,kHz) and on the laser diode injection current in a high bandwidth loop ($\approx 50$\,kHz). Figure~\ref{fig:stabi}(a) shows a typical beat spectrum of the MTS-stabilized ECDL system and a PS-stabilized ECDL system averaged for 16\,s. The beat spectrum has a Gaussian line shape which originates from the convolution of the Gaussian line shapes of both lasers. The beat spectrum has a combined width
\begin{equation}
\sigma_{beat} = \sqrt{\sigma_1^2+\sigma_2^2}
\end{equation}
with the laser linewidths of the two ECDL systems $\sigma_{1,2}$. Pairwise comparison of the three independently stabilized ECDL systems allows us to determine the laser linewidth of each laser. This yields a laser linewidth (FWHM) of $\sigma_{MTS} = (150 \pm 7)\,$kHz for the MTS-stabilized system.

The long-term stability of the MTS-based laser stabilization is recorded by observing the beat note spectrum between the MTS-stabilized and the MSHB-stabilized ECDL systems over 15\,hours. This gives the long-term frequency variation between both lasers. Figure~\ref{fig:stabi}(b) shows a measurement with a maximal frequency deviation of 82\,kHz and a standard deviation of 18\,kHz (rms) which is approximately two orders of magnitude below $\Gamma$. The long-term frequency drift is caused by a change of the ambient temperature. We attribute this to the higher temperature sensitivity of our MSHB setup. Correcting the beat note frequency data for the observed temperature drift reduces the standard deviation to less than 5\,kHz (rms). These results are comparable to other high-performance long-term stabilized laser systems \cite{Zi2017, Torrance2016}. Varying the optical power incident to the MTS setup and observing the center frequency of the beat note spectrum with an independently stabilized ECDL system allows us to determine the intensity-dependent frequency shift of the MTS-stabilized ECDL system. With the alignment-based RAM reduction scheme discussed in Section~\ref{sec:setup}, this shift amounts to less than 20\,kHz within a range in input powers of $\pm 10\%$ of the laser-power operating point. These results show that the effect of RAM is negligible in the present setup. 

\section{Conclusion}

In this work, a universal analysis for determining the optimal modulation parameters for modulation transfer spectroscopy has been presented. The results can be applied to spectroscopy of any atomic species of interest requiring only the knowledge of the effective linewidth $\Gamma_{eff}$. Our analysis shows that optimal signal slopes and amplitudes are obtained for a large modulation index $M$. On the other hand, we also find that the obtainable signal enhancement already saturates for $M \le 10$. For every parameter set, the demodulation phase can be chosen such that the slope or the amplitude are maximized.  In the regime of large $M$ the optimal phase settings for slope and amplitude are very similar. Increasing $M$ shifts the optimal modulation frequency $\omega_m$ to smaller values thus reducing the available bandwidth in a stabilization feedback loop. Taking all these considerations into account, an optimized regime for MTS spectroscopy is given by modulation indices in the range of $3 \le M \le 10$. With this work, the optimal modulation frequency $\omega_m$ can be identified for a chosen $M$ and vice versa. For every parameter set, the value of signal slope and amplitude relative to the optimal values can be determined. An experimental MTS setup using an AOM in double pass configuration is presented and applied to spectroscopy on the ${}^{85}$Rb D2 line. In order to reduce residual amplitude modulation, an alignment procedure for the AOM based on symmetrizing the generated sideband spectrum is introduced. The obtained signal shape is in very good agreement with the theoretical description. The dependence of signal slope and amplitude on $M$ shows only small deviations from the theoretical prediction for a modulation index up to $M = 10$. A close to optimal signal with emphasis on maximizing the slope is obtained for a modulation frequency of $\omega_m = 2\pi\cdot 2.5\,$MHz and a modulation index of $ M = 4.14$. For this set of parameters, 90\% of the maximal slope and 60\% of the maximal amplitude are achieved. Applying the optimized MTS signal, laser stabilization with a laser linewidth of 150\,kHz (averaged for 16\,s) and a long-term stability of 18\,kHz (rms) over 15 hours are demonstrated.

\section*{Funding}
Deutsche Forschungsgemeinschaft (DFG) (Grant No. BI 647/6-1, Priority Program SPP 1929 (GiRyd)); German Research Foundation; Open Access Publishing Fund of Technische Universit\"at Darmstadt.

\bibliography{literatur} 

\end{document}